\documentclass[12pt]{article}

\usepackage[dvips]{graphics}
\usepackage{psfrag}

\begin{document}

\centerline{\bf \large
Geometrical origin of chaoticity in the bouncing ball billiard
}
\vskip 0.5cm

\begin{center}
L.\ M\'aty\'as$^{1}$, and I.\ F.\ Barna$^{2}$
\end{center}

\noindent
$^{1}$ Sapientia University, Department of Technical and Natural Sciences, 
Libert\u{a}tii sq. 1, 
530104 Miercurea Ciuc, Romania. 
\\
$^{2}$  KFKI Atomic Energy Research Institute of the Hungarian Academy of 
Sciences, (AEKI) P.O.\ Box 49, 1525 Budapest, Hungary.

\begin{abstract}
We present a study of the chaotic behavior of the bouncing ball billiard.  
The work is realised on the purpose of finding at least certain 
causes of separation of the neighbouring trajectories. 
Having in view the geometrical construction of the system, 
we report a clear origin of chaoticity of the bouncing ball billiard. 
By this we claim that in case when the floor is made of arc of circles - in 
a certain interval of frequencies -  
a lower bound for the maximal Ljapunov can be evaluated by semianalical 
techniques.  
\end{abstract}

%%************************************************************
\section{Introduction}

Deterministic features of transport has been studied in different  
problems \cite{Do95}. 
These works has also shown that transport may be related 
to the chaotic aspects of the dynamics \cite{ViGa03}.   

The idea of bouncing ball was studied in different problems where analytical 
approximations \cite{Holmes} and comprehensive numerical works can also be 
found \cite{MeLu93}. The analytical approximations have been shown the 
possible evidence of bifurcations while the numerical works 
presented chaotic regimes of the bouncing ball system. 

The {\em bouncing ball billiard} as a spatial extension of the one dimensional 
bouncing ball problem has been introduced in \cite{MaKl04}. 
This first work enhances the irregular diffusivity of the system 
which is similar to certain models of transport \cite{KlKo04}.   
The following work has outlined the spiral modes in the phase space 
of this problem \cite{KlBaMa04} and also pointed out its relevance 
on granular matter \cite{LoCoGo03}. 
Idealised versions where the bounces are performed without loss of energy, 
i.e. the restitution coefficient is one, and there is no oscillation of the 
floor may be found in \cite{HaGa01}.

The {\em chaoticity} of the sawtooth type of the bouncing ball billiard 
has been studied in \cite{WiKa07}, 
with considerable theoretical background \cite{RiBaGeJuKa05}.    
Considering the problem as a gravitational billiard, aspects on chaotic 
features are also approached by numerical methods in \cite{GoSr06}.  

The present work focuses on the geometrical origin of chaoticity 
of the bouncing ball billiard 
where it is investigated the impact of the curvature of the arcs of 
circles on the maximal Ljapunov exponent. The derivation shows, that 
in case of resonance one may give semianalitical 
estimates on chaotic behavior. 
 
The study on manifolds for multi-dimensional billiards related to 
geometric properties is presented in \cite{BaChSzTo03}. 

From the practical point of view the reaction of $CO$ with $O_2$ on 
$Pt$ surface, which is under thermal excitation, 
the molecules $CO$ performs diffusive motion on the 
surface before the reaction would occur \cite{OeRoNe92}.

Quasi-deterministic aspects on diffusion may occur in the behaviour of 
of certain species where in the process of food searching one can find a 
randomness, but there is also a kind of determinism because the 
animals may have certain remembrances on the places where they found 
food in the past \cite{Ma07}.  

The article is organised as follows. In Section \ref{sec:bbb} we 
shortly describe the bouncing ball billiard system.   
The Section \ref{sec:gencons} makes a presentation of 
that frequency region where the semianalical 
approaches to some extent are possible. 
Section \ref{sec:geom} shows an evaluation which has a semi-empirical 
and analytical background giving a lower bound 
for the maximal Ljapunov exponent. Finally, 
Section \ref{sec:thereal} discusses 
the similarities and differences of the analytical and real value of 
the maximal Ljapunov exponent.   

%%*****************************************************************
\section{The bouncing ball billiard }
\label{sec:bbb}

At this point we make a review of the most important features of 
the bouncing ball problem. The system studied is a point particle, 
which bounces on a floor realised of arc of circles \cite{obs}. 
The floor is oscillating with a frequency $f$ corresponding to a circular 
frequency $\omega=2 \pi f$.   
The system is presented below. 
\begin{figure}[h!]
{\hfill
\psfrag{A}{\hspace{-0.1cm}{\Large {$A$}}}
\psfrag{OM}{{\Large {$\omega$}}}
\psfrag{x}{{\Large {$x$}}}
\psfrag{y}{{\Large {$y$}}}
\psfrag{vx}{{\Large {$v_x$}}}
\psfrag{vy}{{\hspace{-0.4cm}\Large {$v_y$}}}
\psfrag{g}{\Huge $g$}
\psfrag{s}{\Large $s$}
\scalebox{0.4}{\includegraphics{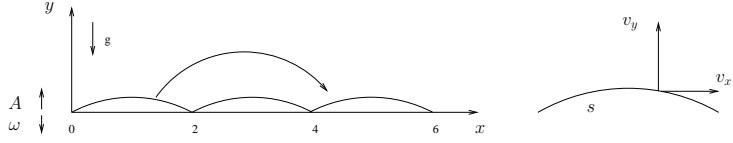}}
\hfill}
\caption[small]{The bouncing ball billiard. 
A point particle realises bounces on a vibrating floor consisting of 
arcs of circles.}
\label{fig:coordinates}
\end{figure}
The bouncing ball billiard from the point of view of the diffusion 
was presented in a comprehensive way in \cite{MaKl04,KlBaMa04}. 
The main conclusions 
are that the system possesses irregular diffusion, and the principal maximas 
for the diffusion occurs at the resonances. These resonances are 
at the frequencies, where the time of the flight becomes equal or multiple of 
the period of vibration applied.

The bouncing ball billiard that we study in this paper, with the floor formed
by circular scatterers, is depicted in Fig.\ \ref{fig:coordinates}.

The equations of motion of this system are presented as follows: The particle
performs a free flight between two collisions in the gravitational field $g \parallel y$. Consequently, its coordinates $(x^-_{n+1},y^-_{n+1})$ and
velocities $(v^-_{x\,n+1},v^-_{y\,n+1})$ at time $t_{n+1}$ immediately before
the $(n+1)^{th}$ collision and its coordinates $(x^+_n,y^+_n)$ and velocities
$(v^+_{x\,n},v^+_{y\,n})$ at time $t_n$ immediately after the $n^{th}$ 
collision
are related by the following equations
\begin{eqnarray} 
x^-_{n+1} &=& x^+_{n} + v^+_{x\,n}(t_{n+1}-t_{n})  \\ 
y^-_{n+1} &=& y^+_{n} + v^+_{y\,n} (t_{n+1}-t_{n}) - g(t_{n+1}-t_{n})^2/2, \\
v^-_{x,\,n+1}&=& v^+_{x\,n}  \\ 
v^-_{y,\,n+1} &=& v^+_{y\,n} - g (t_{n+1}-t_{n})   \:.
\end{eqnarray}
At the collisions the change of the velocities is given by
\begin{eqnarray}
v^+_{\perp \,n}-v_{ci\perp\,n} &=&  k
\,(v_{ci\perp n}-v^{-}_{\perp n})  \\
v^+_{\parallel \, n}-v_{ci\parallel \, n}
&=& 
\beta \,(v^-_{\parallel \, n}-v_{ci\parallel \, n}) \:,
\label{eq:horizrest}
\end{eqnarray} 
where $v_{ci}$ is the velocity of the corrugated floor. 
We distinguish between the two different velocity 
components relative to the normal vector at the 
surface of the scatterers, where the scatterers are represented 
by the arcs of the circles forming the floor.  
$v_{\perp}$, $v_{\parallel}$ and $v_{ci\perp}$, 
$v_{ci\parallel}$ is the normal and tangential  
components of the particle's, respectively the floor's velocity with respect 
to the surface at the scattering point. 
Correspondingly, we introduce two different
restitution coefficients $k$ and $\beta$ that are perpendicular,
respectively tangential to the normal.

As in case of the vertically bouncing ball problem we assume that the floor
oscillates sinusoidally, $y_{ci}=-A\,\sin(\omega\, t)$, where $A$ and $\omega$
are the amplitude respectively the frequency of the vibration, see
Fig.\ref{fig:coordinates}. 

The radius of circles are $R$=15mm and the restitution coefficients 
$k=0.7$, respective $\beta = 0.99$. It is important that 
the slope on the arcs of the circles is very shallow. The distance between 
two arc of circles is $d =  2mm $. 
By this terms proportional with $d^2/R^2$ or less 
are considered terms with second 
- or higher - order. 

%%****************************************************************
\subsection{Considerations on the chaoticity of the bouncing ball}

The chaos of the bouncing ball which spatially is a  
one dimensional system and which is performed on the vertical direction 
has been discussed in \cite{MeLu93}.
 
If we consider $\Delta x_0$ the initial displacement and $\Delta x_n$ the
displacement after $n$ bounces between neighboring trajectories for the one
dimensional vertically bouncing ball, then  
the Ljapunov exponent may be evaluated:  
$\lambda \simeq (1/t_n) \log(\Delta x_n/\Delta x_0)$. 
The main problem is that at certain frequencies the ball may be 
stucked on the surface. 
Because the neigbouring trajectory usually is also stucked, 
i.e. $\Delta x_n$ becomes zero, consequently   
in such cases the Ljapunov exponent $\lambda$ is simply undefined.  

Of coarse one may discuss on chaotic 
regimes between two stucks, but this is the reason why in general 
considerations 
about the chaoticity of the vertically bouncing ball should be done 
in a very careful way.  

In spite of the fact that in two dimensions it is almost impossible 
to have neighboring orbits stucked at the same place, the following 
study will try to avoid that frequency regions where stucking orbits 
might be possible. 
 
%%******************************************************
\section{General considerations}
\label{sec:gencons}

We discuss the chaoticity of the bouncing ball billiard at the frequency
region where the 1/1 resonance holds. 
The approximation we try to make is semi-empirical and the considerations are 
presented below.  
The first observation we make is, that the time of the flight between 
subsequent collision at 1/1 resonance is approximately the same. 
We note this time with $t_{fly}$ and corresponds to that time while 
the particle makes one bounce 
and it is close to the time while the floor 
makes one complete oscillation.  
Subsequent values of $t_{fly}$ at 1/1 resonance are shown below

\begin{tabular}{|c | c|} 
\hline
\mbox{time of flight} & \mbox{collision no}  \\
\hline
0.0183    &  29 \\
0.0191    &  30 \\
0.0185    &  31 \\
0.0190    &  32 \\
0.0184    &  33 \\
\hline
\end{tabular}

One can see that these values do not differ too much. 
They cannot be the same, for
instance because the surface is not flat, but they are close to each other 
and around a specific value.    

The other observation is, that the first resonance manifests so that 
the elongation almost reaches its maximum $A$, and the particle meets the 
floor for almost all cases very close to this height $A$ - see 
fig. \ref{fig:trajectory}. 

%%*****************************************************************
\section{Geometrical origin of chaoticity}
\label{sec:geom}

We consider one component of the velocity, namely the horizontal one $v_x$.  
We are interested in change in difference 
of the horizontal component of the velocities 
of two neighboring trajectories. 
By this we try to make an estimate of a lower bound of 
the Ljapunov exponent, which manifests on the $v_x$ direction of the 
phase space, which would also give a picture on the
{\em horizontal chaoticity} of this problem. 

\begin{figure}[h!] 
{\hfill
\scalebox{0.4}{\includegraphics{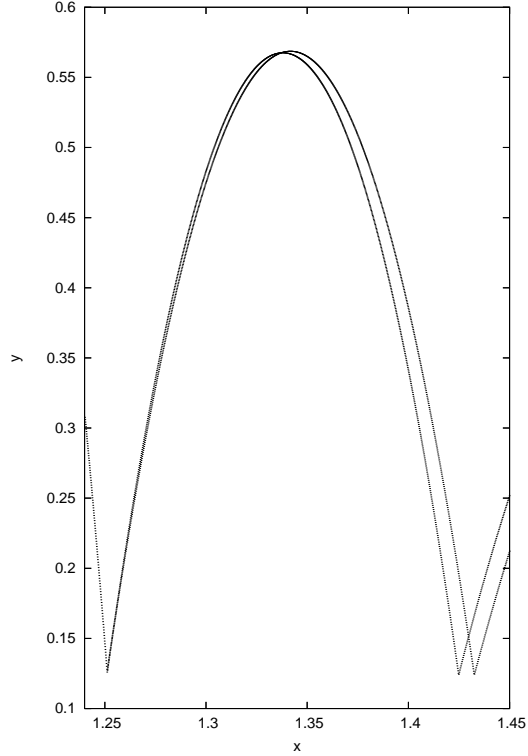}}
\hfill}
\caption[small]{Typical trajectory of the 1/1 resonance. While the particle
arrives at a height close to 0.6, horizontally in general a length around  
0.15 is made. The first conclusion
is, that the particle in most of the cases 
will arrive in a steep angle, so the incident
angle relative to the normal to the surface is small. The other conclusion is,
that quite a number of such bounces occurs while the particle arrives from
one arc to another one.The second trajectory is a neighboring one, caused to 
a deflection of angles of initial velocities at the starting point.}
\label{fig:trajectory}
\end{figure}

The figure below fig.\ \ref{fig:trajectory} 
shows two trajectories, starting from the same place, with slightly different 
velocity vectors.  
The picture is at 1/1 resonance, where after a certain transient 
the bounces are made 
a little bit above the vertical coordinate $A=0.1$ mm which denotes the 
amplitude. 

%%**************************************************************
%%**********************************************************

We assume that at the starting point there is a difference  
in angles but not in the magnitude of the velocities $v_0$. 
This initial deflection in angles we denote by $\delta$. 
The corresponding difference in initial velocities we denote by 
$\Delta v_{x,ini}$.   
Because of this initial deflection in angle there will be at final 
arrival a change in horizontal coordinates $\Delta x$. 

Due to the convex curvature of the arcs characterized 
by the radius $R$, the displacement $\Delta x$ at the arrival on this 
curvature will cause further deflection in angles after one 
collision which we denote by $\delta '$. This cause a 
final difference in the horizontal component of velocities 
after the first bounce $\Delta v_{x,fin}$.   

The rate of exponential separation of the trajectories after such a bounce 
due to the finite radius $R$ 
we denote with $\lambda_{vx,R}$
\begin{equation}
\lambda_{vx,R} 
= 
\frac{1}{t_{fly}} \ln \frac{|\Delta v_{x,fin}|}{|\Delta v_{x,ini}|}
\end{equation}

At the launch of the trajectory we consider 
the magnitude of the velocity $v_0$, 
the angles relative to the vertical  
are $\alpha^{(1)}$ and $\alpha^{(2)}$, the 
complementer angles are $\alpha_c^{(1)}$ and $\alpha_c^{(2)}$ where 
as it was mentioned above $\alpha^{(2)}=\alpha^{(1)}+\delta$  
\begin{equation}
|\Delta v_{x,ini}| = |v_0 \cos \alpha_c^{(1)} -v_0 \cos \alpha_c^{(2)}|  
                = |v_0 \sin \alpha^{(1)} -v_0 \sin \alpha^{(2)}| 
                = |v_0 \, \delta \cos \alpha^{(1)} | + h.o.t  
\end{equation} 
where the higher order terms means terms which are proportional with at least 
the second power of $\delta$.  
Correspondingly if $\delta '$ is the angle between the directions 
of trajectories after the first bounce 
\begin{equation}
|\Delta v_{x,fin}| = |v_0 \, \delta ' \cos \alpha^{(1)'} | + h.o.t 
\end{equation}
Because there is a free flight in gravitational field and  
the arcs are with shallow slope practically 
$\alpha^{(1)'}\simeq\alpha^{(1)}$, very precisely their difference is 
a second order term.  We mention, that $\alpha^{(1)}$ is also small 
as one can see on the fig.\ \ref{fig:trajectory}, so 
its product with $\delta$ is also considered a value with second order. 
\footnote{Even if a term proportional with 
$\alpha$ or $\sin \alpha$ would be kept, 
at the end where the average value is calculated for $\lambda_{vx}$ or it drops
out or it is proved to be of higher order.}

As a result we get for the value $\lambda_{vx,R}$, which has a definite 
contribution due to the finite value of 
\begin{equation}
\lambda_{vx,R} = \frac{1}{t_{fly}} \ln \left| \frac{\delta '}{\delta} \right| 
+ h.o.t.
\label{eq:lambda3}
\end{equation}
where one can see that the ratio between the deflection 
of the angles after the bounce and before the bounce counts. 

%%%!!!!!!!!!!!!!!!!!!!!!!!!!!!!!!!!!!!!!!!!

\begin{figure}[h!]
\psfrag{O1}{\Large $O_1$}
\psfrag{O2}{\Large $O_2$}
\psfrag{A1}{\Large $\gamma_1'$}
\psfrag{A2}{\Large $\gamma_2'$} 
\psfrag{A2C}{\Large $\alpha_c^{(2)}$}
\psfrag{TH}{\Large $\theta$}
\psfrag{dTH}{\Large $d\theta$}
\psfrag{R}{\Large $R$}
{\hfill
\scalebox{0.4}{\includegraphics{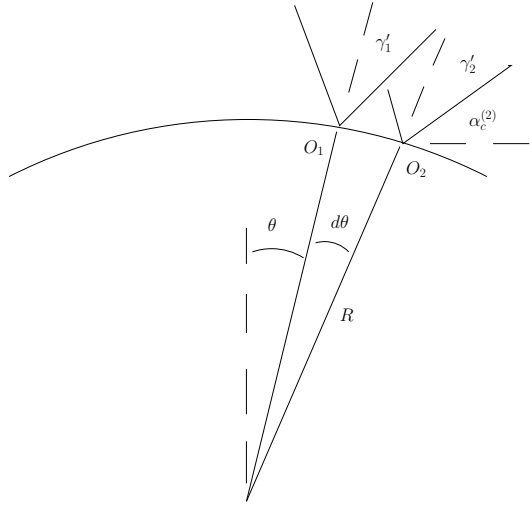}} 
\hfill}
\caption[small]{Illustration of the bounce of neighboring 
trajectories on an arc of circle. The figure enhances the displacement  
of trajectories and their velocities 
after the bounce due to the geometry of the floor.}
\label{fig:arrival}
\end{figure}

The bounce is presented on 
Fig.\ \ref{fig:arrival}. 

Based on Fig.\ \ref{fig:arrival} we can conclude that the deflection 
between the two reflected trajectories 
($\delta '$) one hand is due to the initial 
deflection $\delta$. 
For the reflected trajectories 
there is a contribution due to the curvature. 
If the point of incidence of the first trajectory is at $\theta$ then    
the point of incidence of the second one is at $\theta+d\theta$. 
The incident angles relative to the normal differs by $d\theta$  
and in addition the reflected angles - on the Fig.\ 
\ref{fig:arrival} - $\gamma_1 '$ and $\gamma_2 '$ has also a difference  
$d\theta$. Consequently the term $d\theta$ has to be counted twice.  
\begin{equation}
\delta ' = \delta+ 2 d\theta
\end{equation}  
Inserting this relation in eq.(\ref{eq:lambda3}) one gets
\begin{equation} 
\lambda_{vx,R} = 
\frac{1}{t_{fly}} \ln \left| \frac{\delta+2d\theta}{\delta} \right| + h.o.t.
=\frac{1}{t_{fly}} \ln \left| 1+2\frac{d\theta}{\delta} \right|
+ h.o.t.
\end{equation}

At the arriving point on the surface the initial difference $\delta$ will cause 
a displacement $\Delta x$. This means that on the arc of the circle the
trajectories will arrive at a difference $d\theta \simeq |\Delta x|/R$ 
as one can see on fig.\ \ref{fig:arrival}.  
Now follows the evaluation of $\Delta x$ and correspondingly of the $d\theta$.

The length $\Delta x$ is due to the difference in angles of the velocities, 
at the starting point. The first one is launched at an angle $\alpha_c^{(1)}$ 
the other one with an angle $\alpha_c^{(2)}=\alpha_c^{(1)}-\delta$.  
At the end one of the particles arrives at $x_2$, the other one at $x_1$
\begin{equation}
\Delta x = x_1-x_2  
= 2 \frac{v_0^2}{g} 
[\cos \alpha_c^{(1)} \sin \alpha_c^{(1)}-\cos \alpha_c^{(2)} \sin \alpha_c^{(2)}]
\end{equation}
where $\alpha_c^{(1)}$ is the angle of the velocity made with the horizontal 
direction of the first trajectory at the starting point. 
During the evaluation we make the approximation, that $\sin \delta$ is 
approximately $\delta$ and $\cos \delta \simeq 1$ or 
the differences are at least second order in $\delta$, and are included in 
the higher order terms. 
\begin{equation}
\Delta x = x_1 - x_2 =
2 \frac{v_0^{2}}{g} \delta 
[\cos^2 \alpha_c^{(1)} - \sin^2 \alpha_c^{(1)} ] + h.o.t.  
\end{equation}
At this point one can see that $ \cos^2 \alpha_c^{(1)}=\sin^2 \alpha^{(1)}$ 
- here $\alpha^{(1)}$ being the incident angle relative to the vertical -  
can be neglected, consequently 
$d\theta$ yields the following value  
\begin{equation}
d\theta\simeq\frac{|\Delta x|}{R} 
\simeq 2 \delta \frac{v_0^2}{gR} ( \sin^2 \alpha_c^{(1)} ) 
\end{equation}
If we take into account that the time for the flight 
is $t_{fly}=2 v_0 \sin \alpha_c^{(1)}/g$ then 
\begin{equation}
2\frac{d\theta}{\delta} 
\simeq 4  \frac{v_0^2}{gR} ( \sin^2 \alpha_c^{(1)} )
\simeq \frac{g t^2_{fly}}{R}
\end{equation}
By this we get for the value $\lambda_{vx,R}$ in leading order 
\begin{equation}
\lambda_{vx,R} = \frac{1}{t_{fly}} 
 \ln  ( 1+\frac{g t^2_{fly}}{R} ) + h.o.t. 
= \frac{g t_{fly}}{R} + h.o.t
\end{equation}  
where the logarithm has been expanded, and terms proportional with 
$1/R^2$ have been also considered as being of higher order. 
 
The average of $\lambda_{vx,R}$ means averaging the expression 
above. By this the higher order terms vanishes or becomes 
smaller so they remain of higher order. 
Consequently it reduces to the average of the time of the flight $t_{fly}$. 
Its average is given by the period of oscillations $T$ resulting   
for $\bar{\lambda}_{vx,R}$ in leading order 
\begin{equation}
\label{eq:lambdavx}
\bar{\lambda}_{vx,R} \simeq \frac{g T}{R} \simeq \frac{gK}{f} 
\end{equation}
Because $\bar{\lambda}_{vx,R}$ is a manifestation of the separation of the 
neighboring trajectories in the $v_x$ direction due to the geometry, 
consequently this value 
can be considered a lower bound for the maximal Ljapunov exponent.   

%%*************************************************************
\subsection{The case of two periodic orbits} 

As it is pointed out in the work \cite{Holmes} - 
with increasing the frequency - bifurcation of the resonant 
trajectory may be possible. This means that the time of 
flight consists of a shorter and a longer time alternating 
one after the other which we denote by $t_{fly,1}$ and $t_{fly,2}$, 
but they still do not differ too much from each other.  
\footnote{
In general in the case of the bouncing ball billiard to have a considerable 
difference between $t_{fly,1}$ and $t_{fly,2}$ even it is not possible, 
or because the dynamics enters in 
further bifurcations, or simply it crashes to a scenario with lots of sticking 
orbits.}

In such case the approximation that have been presented previously 
are still valid and one gets after two consequent 
flights - one is shorter, one is longer - 
\begin{equation}
\lambda_{vx,R} = \frac{1}{t_{fly,1}+t_{fly,2}} 
 \ln  \left| \frac{\delta'}{\delta}\frac{\delta''}{\delta'} \right|  
+ h.o.t. 
\end{equation}  
The argument of the logarithm can be written as 
\begin{equation}
\lambda_{vx,R} = \frac{1}{t_{fly,1}+t_{fly,2}}
\ln  \left[ \left( 1+\frac{g t^2_{fly,1}}{R} \right) \,
    \left( 1 + \frac{t^2_{fly,2}}{R} \right) \right] + h.o.t. 
\end{equation}
After the expansion to the first order we have 
\begin{equation}
\lambda_{vx,R} = \frac{g}{R}
\left(
\frac{t^2_{fly,1}+t^2_{fly,2}}{t_{fly,1}+t_{fly,2}} 
\right) + h.o.t.
\end{equation}
In the numerator of the second fraction 
the decompositions are made 
\begin{equation}
t_{fly,1(2)}=\frac{t_{fly,1}+t_{fly,2}}{2}\pm\frac{t_{fly,1}-t_{fly,2}}{2} 
\end{equation}
Finally we arrive to the relation 
\begin{equation}
\lambda_{vx,R} = \frac{g}{R}
\frac{t_{fly,1}+t_{fly,2}}{2} 
+  \frac{g}{R}
\frac{(t_{fly,1}-t_{fly,2})^2}{2(t_{fly,1}+t_{fly,2})} + h.o.t.
\end{equation}
This expression can be averaged and the term proportional with 
$(t_{fly,1}-t_{fly,2})^2$
is still too small and is considered of second order.
The average of the last expression yields the value    
\begin{equation} 
\bar{\lambda}_{vx,R} \simeq \frac{g}{R} 
\frac{(T_1+T_2)}{2}
\end{equation} 
where $T_1$ and $T_2$ represents the average values of the 
$t_{fly,1}$ respective $t_{fly,2}$. 
Because even in the case of 
the two periodic orbit $(T_1+T_2)/2$ equals an average time flight $T'$ 
which is the inverse of 
that frequency $f$ where the dynamics is already bifurcated. 
This is in fact an interval of frequencies, so 
the latter formula for two periodic orbits 
still shows a strong analogy with 
eq.\ (\ref{eq:lambdavx}) and it may be written 
\begin{equation} 
\bar{\lambda}_{vx,R} \simeq \frac{g}{R f} 
\end{equation} 
This latter formula shows 
that the relation (\ref{eq:lambdavx})
may be valid for the full 
1/1 resonance and for bifurcated trajectories not too far from it.  

%%****************************************************************
\section{The real value of the maximal Ljapunov exponent}
\label{sec:thereal}

In this section we discusses connections of the geometry with 
the chaoticity.
The relation (\ref{eq:lambdavx}) {\em in case of the 1/1 resonance} 
at a given gravitational 
field $g$ has a characteristic time of flight $T$. 
Because of this reason, one may consider that one of the most 
relevant dependence of the maximal Ljapunov exponent 
from practical purposes is in terms of the radius. 
So in this section we restrict ourselves to the case of the frequency 
$f=53.2$ Hz when the first resonance is fully developed. 
\footnote{
The interval where we expect a validity of the relation 
(\ref{eq:lambdavx}) is a frequency region $f=53.2\pm 5$ Hz. 
The validity sometimes may be even wider, however there is 
a slight dependence of the endpoints of the interval on $R$, 
consequently it has been chosen a common domain in which 
the resonance or eventually the bifurcated scenario holds for different 
radii we discuss below.   
}   
The following figure illustrates the numerical values of the 
maximal Ljapunov exponent in terms of inverse of the radius. 
Consequently we start at a curvature 1/24 mm$^{-1}$ 
and we end at a curvature 1/16 mm$^{-1}$. 
The endpoints are also motivated by the works \cite{MaKl04,KlBaMa04}. 
%%*************************************************
\begin{figure}[h!]
{\hfill
\scalebox{0.4}{\rotatebox{270}{\includegraphics{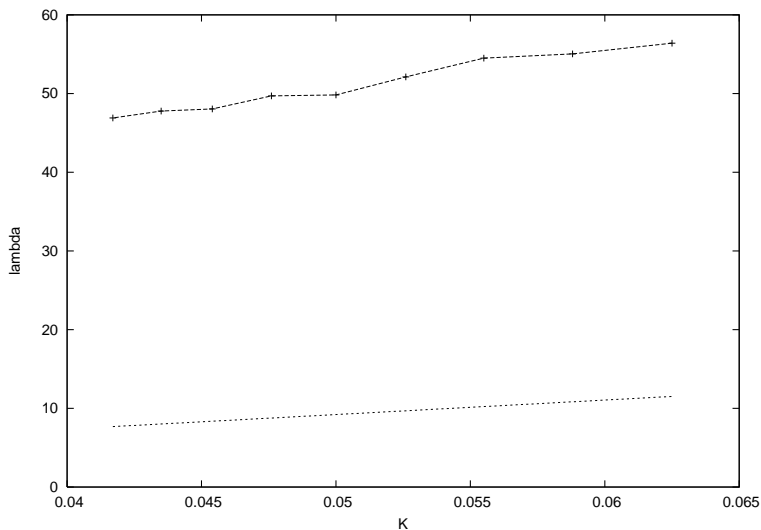}}} 
\hfill}
\caption[small]{Presentation of the maximal Ljapunov exponent 
as a function of the inverse of the radius,\ i.e. the curvature $K$, 
at the frequency $f=53.2$ Hz. 
The continuous straight line presents the lower bound
estimation (\ref{eq:lambdavx}). 
}
\label{fig:Ljapexp}
\end{figure}
%%*************************************************

As one can see the maximal Ljapunov exponent increases as the 
curvature - the inverse of the radius - increases.  
This is normal because a smaller radius implies a stronger 
separation of neighboring trajectories. 
From quantitative point of view  
both the numerical and the analytical evaluations
has similar behavior - as it is shown on the fig.\ \ref{fig:Ljapexp}.

Regarding the quantitative aspects the evaluation (\ref{eq:lambdavx}) 
is considerably below of the real value. Partly because the analytical 
evaluation is just a projection. 
On the other hand the evaluation wants to detect only the geometrical 
effects on the chaoticity. Of coarse considerable other effects may have 
an important role on the chaoticity, but the study of such aspects will 
need further work.   
 
We thank Rainer Klages for comments and encouragement.    
%%*************************************************************

%%*****************************************************************


\begin{thebibliography}{xx}
\bibitem{obs} 
One can also study the bounce of a particle with finite radius, that can 
be reduced to a point particle problem \cite{MaKl04}
\end{thebibliography}
\end{document}